\documentclass[twoside,twocolumn,9pt]{article}
\usepackage{extsizes}
\usepackage[super,sort&compress,comma]{natbib} 
\usepackage[version=3]{mhchem}
\usepackage[left=1.5cm, right=1.5cm, top=1.785cm, bottom=2.0cm]{geometry}
\usepackage{balance}
\usepackage{mathptmx}
\usepackage{sectsty}
\usepackage{graphicx} 
\usepackage{lastpage}
\usepackage[format=plain,justification=justified,singlelinecheck=false,font={stretch=1.125,small,sf},labelfont=bf,labelsep=space]{caption}
\usepackage{float}
\usepackage{fancyhdr}
\usepackage{fnpos}
\usepackage[english]{babel}
\addto{\captionsenglish}{%
  
}
\usepackage{array}
\usepackage{droidsans}
\usepackage{charter}
\usepackage[T1]{fontenc}
\usepackage[usenames,dvipsnames]{xcolor}
\usepackage{setspace}
\usepackage[compact]{titlesec}
\usepackage{hyperref}

\usepackage{epstopdf}

\definecolor{cream}{RGB}{222,217,201}

\begin{document}


\makeFNbottom
\makeatletter
\renewcommand\LARGE{\@setfontsize\LARGE{15pt}{17}}
\renewcommand\Large{\@setfontsize\Large{12pt}{14}}
\renewcommand\large{\@setfontsize\large{10pt}{12}}
\renewcommand\footnotesize{\@setfontsize\footnotesize{7pt}{10}}
\makeatother

\renewcommand{\thefootnote}{\fnsymbol{footnote}}
\renewcommand\footnoterule{\vspace*{1pt}%
\color{cream}\hrule width 3.5in height 0.4pt \color{black}\vspace*{5pt}} 
\setcounter{secnumdepth}{5}

\makeatletter 
\renewcommand\@biblabel[1]{#1}            
\renewcommand\@makefntext[1]%
{\noindent\makebox[0pt][r]{\@thefnmark\,}#1}
\makeatother 
\renewcommand{\figurename}{\small{Fig.}~}
\sectionfont{\sffamily\Large}
\subsectionfont{\normalsize}
\subsubsectionfont{\bf}
\setstretch{1.125} 
\setlength{\skip\footins}{0.8cm}
\setlength{\footnotesep}{0.25cm}
\setlength{\jot}{10pt}
\titlespacing*{\section}{0pt}{4pt}{4pt}
\titlespacing*{\subsection}{0pt}{15pt}{1pt}

\fancyfoot{}
\fancyfoot[LO,RE]{\vspace{-7.1pt}\includegraphics[height=9pt]{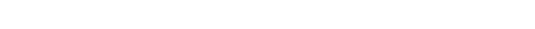}}
\fancyfoot[CO]{\vspace{-7.1pt}\hspace{11.9cm}\includegraphics{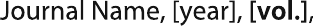}}
\fancyfoot[CE]{\vspace{-7.2pt}\hspace{-13.2cm}\includegraphics{head_foot/RF}}
\fancyfoot[RO]{\footnotesize{\sffamily{1--\pageref{LastPage} ~\textbar  \hspace{2pt}\thepage}}}
\fancyfoot[LE]{\footnotesize{\sffamily{\thepage~\textbar\hspace{4.65cm} 1--\pageref{LastPage}}}}
\fancyhead{}
\renewcommand{\headrulewidth}{0pt} 
\renewcommand{\footrulewidth}{0pt}
\setlength{\arrayrulewidth}{1pt}
\setlength{\columnsep}{6.5mm}
\setlength\bibsep{1pt}

\makeatletter 
\newlength{\figrulesep} 
\setlength{\figrulesep}{0.5\textfloatsep} 

\newcommand{\topfigrule}{\vspace*{-1pt}%
\noindent{\color{cream}\rule[-\figrulesep]{\columnwidth}{1.5pt}} }

\newcommand{\botfigrule}{\vspace*{-2pt}%
\noindent{\color{cream}\rule[\figrulesep]{\columnwidth}{1.5pt}} }

\newcommand{\dblfigrule}{\vspace*{-1pt}%
\noindent{\color{cream}\rule[-\figrulesep]{\textwidth}{1.5pt}} }

\newcommand*{\DCH}{$K^{-1}K^{-1}V^{2}$ }
\newcommand*{\SCH}{$K^{-1}V^{1}$ }

\makeatother

\twocolumn[
  \begin{@twocolumnfalse}
\sffamily

\noindent\LARGE{\textbf{Resonant Double-Core Excitations with Ultrafast, Intense  Pulses}} \\

\noindent\large{Adam E. A. Fouda,$^{\ast}$\textit{$^{a\dag}$} Dimitris Koulentianos,\textit{$^{\dag\ddag}$}, Linda Young{$^{\dag\P}$}, Gilles Doumy$^{\ast}$\textit{$^{b\dag}$} and Phay J. Ho$^{\ast}$\textit{$^{c\dag}$}} \\

\includegraphics{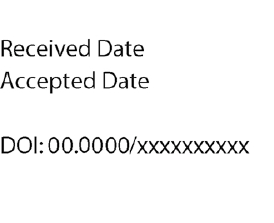} \noindent\normalsize{

Intense few-to-sub-femtosecond soft x-ray pulses can produce neutral, two-site excited double-core-hole states by promoting two core electrons to the same unoccupied molecular orbital. We theoretically investigate  double nitrogen K-edge excitations of nitrous oxide (N$_2$O) with multiconfigurational electronic structure calculations. We show that 
the second core-excitation energy is reduced with respect to its ground state value. A site-selective double core-excitation mechanism using intense few-femtosecond x-rays is investigated using time-dependent Schrodinger equation (TDSE) simulations. The subsequent two-step Auger-Meitner and two-electrons-one-electron decay spectra of the double core-excited states are analyzed using a Mulliken population analysis of the multiconfirational wavefunctions. The change in the electron emission lineshape between the absorption of 1 or 2 photons in the resonant core-excitation is predicted by combining this approach with the TDSE simulations. We examine the possibility of resolving the double core-excited states with x-ray pump-probe techniques by calculating the chemical shifts of the core-electron binding energy of the core-excited states and decay products.} \\

 \end{@twocolumnfalse} \vspace{0.6cm}

  ]

\renewcommand*\rmdefault{bch}\normalfont\upshape
\rmfamily
\section*{}
\vspace{-1cm}


\footnotetext{\textit{$^{a}$ E-mail: foudaae@anl.gov}}
\footnotetext{\textit{$^{b}$E-mail: gdoumy@anl.gov }}
\footnotetext{\textit{$^{b}$E-mail: pho@anl.gov }}
\footnotetext{\dag~Chemical Sciences and Engineering Division, Argonne National Laboratory, 9700 S Cass Ave, Lemont, IL 60439, USA.}

\footnotetext{\ddag~Present Address: Center for Free-Electron Laser Science, Deutsches Elektronen-Synchrotron DESY, Notkestra{\ss}e 85, 22607 Hamburg, Germany \& Department of Physics, Universit\"at Hamburg, Luruper Chaussee 149, 22761 Hamburg, Germany}
\footnotetext{\P~Department of Physics and James Franck Institute, The University of Chicago,
Chicago, Illinois 60637, USA}


\section{Introduction}

The resonant interaction of high-intensity x-ray free electron laser (XFEL) pulses with atoms and molecules on few-to-sub-femtosecond timescales holds promise to a variety of new applications; such as selectively exciting specific local electronic states through the stimulated x-ray Raman process,\cite{mukamel2013multidimensional,rohringer2019x,o2020electronic} enhancing scattering signal images without significant x-ray damage \cite{kanter2011unveiling,rudek2013resonance,ho2014theoretical,ho2020role} and producing coherent superpositions of core-excited states\cite{li2022attosecond}. High intensity XFEL experiments have the potential to induce low probability processes such as valence photoionization and and multi-photon absorption.
With intense attosecond pulses, the possibility of sequential absorption before decay processes can be high, and it could significantly impact the efficiency of the stimulated x-ray Raman scattering and enhanced scattering signals. 

Early experiments at the Linac Coherent Light Source (LCLS) XFEL used intense femtosecond x-ray pulses for the experimental observation of molecular double-core-hole (DCH) states\cite{berrah2011double,salen2012experimental,koulentianos2020high}, produced through a sequential multi-photon absorption mechanism that creates two vacancies either on a single atomic site (ss) or on two sites (ts) of a molecule. Such states were initially described in a seminal theory work by Cederbaum \textit{et al.}\cite{cederbaum1986double}. Synchrotron radiation (SR) can also produce ss and ts DCH states with low probability by relying on the change in dynamic-electron correlation induced by the sudden removal of a core electron\cite{eland2010double,lablanquie2011properties,lablanquie2011evidence,nakano2013single,carniato2019double}, as well as more exotic DCH states through a core-ionization or core-excitation mechanism\cite{carniato2015single1,carniato2015single2,PhysRevLett.111.123001,PhysRevLett.114.093001,carniato2016photon,PhysRevLett.117.133001,feifel2017cationic,koulentianos2018kl,koulentianos2018double,marchenko2020single,koulentianos2021ok} and recently these states were observed using an XFEL\cite{mazza2020mapping}. However, to the best of our knowledge, the formation of neutral excited ss or ts-K- shell DCH states created through sequential core-excitations is unobserved.  We note that KL excitations of this form (involving $s$ and $p$-shells) have been observed for argon and the sulfur-containing molecules H$_2$S, SO$_2$ and SF$_6$\cite{vstuhec1994configuration,reynaud1996double} in SR experiments. 
\begin{figure*}[h]
\centering
  \includegraphics[scale=0.8]{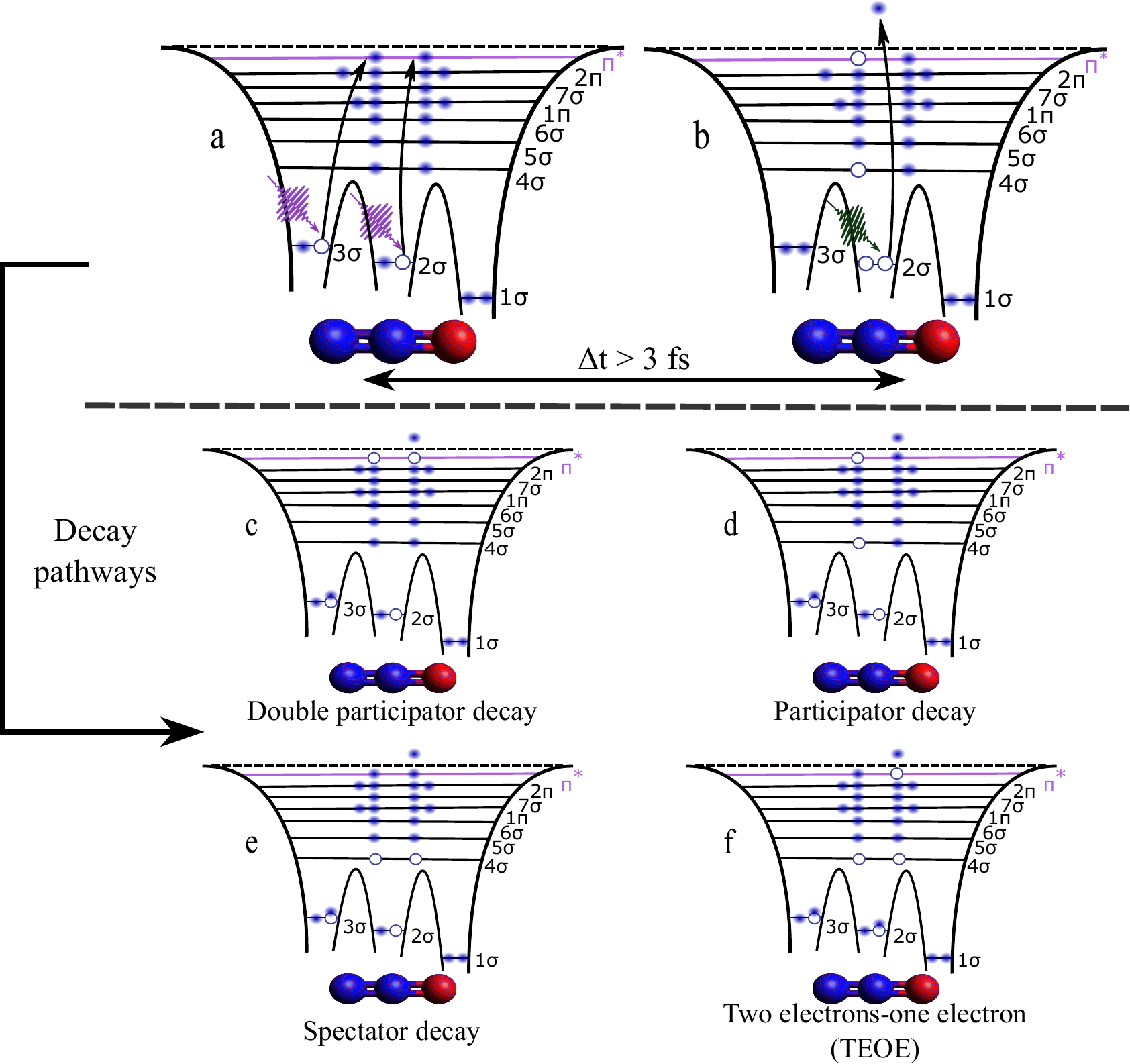}
  \caption{a) Double K-shell excitation from the two nitrogen sites to the $\pi^{*}$ orbital. b) Probing the DCH decay with time resolved-x-ray photoelectron spectroscopy (TR-XPS). c) Double participator decay involving both core-excited electrons. d) Participator process involving both an inner-shell electron and the core-excited electron. e) Spectator process involving two inner-shell electrons. f) Two-eletrons-one-electron (TEOE) process involving two electrons simultaneously filling both core-holes and ejecting a single electron.}
  \label{fgr:schematic}
\end{figure*}

Nitrous oxide (N$_{2}$O) is a model system for studying neutral excited ts-DCH states, since it is a small molecule with two distinct nitrogen atomic sites with a significant chemical shift in the core electron binding energy (CEBE) of the two nitrogen atoms around 4 eV. Previous SR studies demonstrate selective excitation of the two nitrogen sites  due to the large difference of their binding energy \cite{murphy1988site,lebrun1993dissociation,bozek1993asymmetry,ferrand1996new,miron2001nuclear,yu2003anionic,ceolin2008study,knie2014angle,deinert2020interatomic}. Following resonant excitation, the N$_{2}$O Auger-Meitner decays, where one electron is emitted and another fills the core hole. The high-resolution de-excitation spectrum (often referred to as resonant Auger-Meitner spectroscopy) has been measured following single excitations and reveals significantly different electronic kinetic energy distributions depending on the excitation site\cite{larkins1988core,piancastelli2007high}. A clear signature of the decay of core-excited molecules is from the so-called participator channels, where the electron excited in the first step also participates in the subsequent decay. Because the final state of this decay process is the same as the one following direct valence photoionization of the ground state molecule, the kinetic energy of these electrons can be easily predicted. 

Fig. \ref{fgr:schematic} a) shows the generation of a doubly core-excited N$_{2}$O molecule via ts double nitrogen K-shell excitation to the $\pi^{*}$ orbital. Scenarios for detecting this state are shown, b) using time-resolved x-ray photoelectron spectroscopy (TR-XPS) and c-f) through tracking the variety of decay pathways. 
The double-participator decay in c) is unique to these states and we will show that its signature is of higher kinetic energy than other pathways. d) and e) show the participator and spectator decay processes which are analogous to single-core excited state decay. One intriguing possibility is observing the two-electron-one-electron (TEOE) process shown in f), where both core holes are filled simultaneously and all the excess energy is carried by a single continuum electron of much higher energy. This process has only been observed previously using SR, for inner-shell vacancies in argon\cite{vzitnik2016two} and the CH$_3$F molecule\cite{feifel2016ultrafast}. It was found that the TEOE decay in argon was 10${^3}$ times less likely to occur than the sequential Auger-Meitner decay\cite{vzitnik2016two}.

In this work we theoretically investigate the production of ts double core excited states (\DCH) through sequential inner shell excitations (via single core excited states (\SCH)) at the nitrogen K-edge in nitrous oxide and characterize the states via the resonant Auger decay. Sections \ref{sec:methods} and \ref{sec:results} present the calculation details and results discussion, which are both split into four complementary subsections. First, x-ray absorption near edge structure (XANES) spectra detailing the required energetics for the two single-photon excitation yielding the \DCH states are discussed in Subsections \ref{sec:xanescalc} and \ref{sec:xanes}. Subsections \ref{sec:tdsecalc} and \ref{sec:tdse} discuss the electronic population transfer dynamics in the presence of ultrafast x-rays producing the \DCH states, modelled by numerically solving the time-dependent Schrodinger equation (TDSE)\cite{rohringer2007x,rohringer2008resonant,cavaletto2012resonance,li2020resonant,picon2017time,picon2018auger,oberli2019site,fouda2021site}. Subsections \ref{sec:augercalc} and \ref{sec:auger} discuss the resonant Auger-Meitner decay of the  \DCH states, which undergo two-step decay and TEOE processes. Finally, chemical shifts in the core-electron binding energy (CEBE) for the core-excited states and the associated Auger-Meitner decay products are discussed in Subsections \ref{sec:cebecalc} and \ref{sec:cebe} respectively.

\section{Theoretical and Computational Details}\label{sec:methods}

The study uses the restricted active space self consistent field (RASSCF) multiconfigurational quantum chemical method\cite{malmqvist1990restricted,werner1981quadratically} and its second order perturbation correction to correlation energy (RASPT2)\cite{malmqvist2008restricted,finley1998multi}. There are numerous approaches for calculating core-excited states from first principles\cite{zhang2014nonlinear,norman2018simulating,besley2020density,https://doi.org/10.1002/wcms.1527,rankine2021progress}. Here we use RASSCF as it is a widely used approach for core-hole spectroscopy and provides a flexible and accurate approach to treating single and multiple core-hole states in small gas-phase molecules and transition metal complexes\cite{koulentianos2020high,delcey2019efficient,vacher2020origin,C9CP03019K,fouda2020observation}. All electron structure calculations in this work were calculated using the OpenMolcas software package\cite{fdez2019openmolcas}. A fixed ground state geometry was used throughout, optimised by density functional theory (DFT) with the B3LYP exchange correlation functional and the 6-311G$^{**}$ basis set. The resulting N$_{c}$-N$_{t}$ and N$_{c}$-O bond-lengths were 1.12 and 1.18 \AA. Unless stated otherwise, all calculations in this work were calculated with the ANO-RCC-VQZP basis set in combination with the Cholesky-based resolution of the identity auxiliary basis\cite{aquilante2008accurate}. No point group symmetry restrictions were applied in this work.

\subsection{XANES Spectra}\label{sec:xanescalc}
The ground state, two N$_{t}$ and N$_{c}$ \SCH states (1$s\rightarrow\pi^{*}_{x}$ and 1$s\rightarrow\pi^{*}_{y}$) and four \DCH states (where K is the N 1s orbitals and V is the doubly degenerate $\pi^{*}$) were initially calculated at the state-averaged (SA-)RASSCF(20,1,2;1,11,1) level of theory. This uses the RASSCF($n$,$l$,$m$;$i$,$j$,$k$) notation, where $n$, $l$ and $m$ refer to the total number of active electrons, the maximum number of holes allowed in RAS1 and the maximum number of electrons allowed in RAS3 respectively; $i$, $j$, and $k$ refer to the number of orbitals in the RAS1, RAS2 and RAS3 subspaces respectively.  Both RAS1 and RAS3 contain either the N$_{t}$ or N$_{c}$ 1s orbitals and the highly excited state (HEXS) scheme\cite{delcey2019efficient}, was applied to both RAS1 and RAS3 to generate \DCH states. For \SCH states, HEXS was applied to either RAS1 or RAS3. Throughout this work, the oxygen $1s$ orbital (or any core orbital not included in RAS1 or RAS3) was optimized in the Hartree-Fock (HF) level of theory, and the remaining occupied orbitals were in RAS2. The unoccupied $\pi_{x}^{*}$, $\pi_{y}^{*}$ and Rydberg $3s$ orbitals were also included in RAS2. Dynamic correlation was included via multi-state restricted active space second order purtbuation theory (MS-RASPT2)\cite{malmqvist2008restricted,finley1998multi} with an imaginary shift of 0.1 a.u. The transition dipole moments between the ground to \SCH and \SCH to \DCH states were calculated using the restricted active space state interaction (RASSI) method\cite{malmqvist2002restricted}. A broadening of 0.11 eV\cite{miron2001nuclear} and 0.22 eV was applied to the Gaussian lineshapes of the \SCH and \DCH spectra, which assumes the latter decay twice as fast.

\subsection{Time-Dependent Schrodinger Equation}\label{sec:tdsecalc}
The TDSE simulations of the population transfer dynamics of the \DCH states were adopted from the procedure previously used in the simulation of stimulated x-ray Raman scattering (SXRS) on NO\cite{o2020electronic,fouda2021site}. The time-dependent electronic Hamiltonian, $\hat{H}(t)$) in the presence of an external electric field is given by
\begin{equation}\label{eq:hamil}
    \hat{H}(t) = \hat{H}_{0} + V(t),
\end{equation}
where $\hat{H}_{0}$ is field-free electronic Hamiltonian and $V(t)$ describes the interaction between the molecule and the external field. We solve the time-dependent Schr\"{o}dinger equation,
\begin{equation}\label{eq:tdse}    
    i\frac{\partial}{\partial(t)}|\Psi,t\rangle = \hat{H}(t)|\Psi,t\rangle 
\end{equation}
for a set of electronic states determined by the state vector, 
\begin{equation}\label{eq:wf}
    |\Psi,t\rangle = c_{g}(t)e^{-iE_{g}t}|g\rangle + \sum_{s=1}^{N_{s}}c_{s}(t)e^{-iE_{s}t}|s\rangle + \sum_{d=1}^{N_{d}}c_{d}(t)e^{-iE_{d}t}|d\rangle .
\end{equation}
In this ansatz for the wavefunction, the ground state is indicated by $|g\rangle$, the single core excited \SCH states indicated by $|s\rangle$ and the double core excited \DCH states indicated by $|d\rangle$. $E_{g}$, $E_{s}$ and $E_{d}$ are the energies of the ground, \SCH and \DCH states respectively. $N_{s}$ and $N_{d}$ are the total numbers of the \SCH and \DCH states respectively. 
The respective equations of motion (EOM) for the ground, \SCH and \DCH states are
\begin{equation}\label{eq:eomg}
\begin{split}
        i\dot{c}_{g}(t) =\big( E_{g} - \frac{ i\Gamma_{g}(t)}{2}\big)c_{g}(t) &+ \sum_{s=1}^{N_{s}}\gamma^{s}_{g}(t) c_{s}(t)
\end{split}
\end{equation}
\begin{equation}\label{eq:eoms}
\begin{split}
        i\dot{c}_{s}(t) =\big( E_{s} - \frac{ i\Gamma_{s}(t)}{2}\big)c_{s}(t) &+ 
        \gamma^{g}_{s}c_{g}(t) + \sum_{d=1}^{N_{d}}\gamma^{d}_{s}(t) c_{d}(t)
\end{split}
\end{equation}
\begin{equation}\label{eq:eomd}
\begin{split}
        i\dot{c}_{d}(t) =\big( E_{d} - \frac{i\Gamma_{d}(t)}{2}\big)c_{d}(t) &+ \sum_{s=1}^{N_{s}}\gamma^{s}_{d}(t) c_{s}(t).
\end{split}
\end{equation}
The first terms are the diagonal matrix elements, containing the energy $E_{n}$ of the state and its decay width $\Gamma_{n}$. Here the index $n$ is used for terms applying to either $g$, $s$ and $d$. Unless stated otherwise, all quantities and formulas are expressed in atomic units throughout this section. The second terms are the off-diagonal matrix elements, involving the interaction terms $\gamma^{i}_{j}(t)$, which couple the molecular system to the x-ray field within the dipole approximation, 
\begin{equation}\label{eq:interact}
    \gamma^{i}_{k}(t) = D^{\boldsymbol{\mu}}_{ij}\mathcal{E}(t)\cos(\omega_{x}t),
\end{equation}
\begin{equation}\label{eq:gaussian}
    \mathcal{E}(t) = E_{max}e^{\frac{-(t - t_{max})^{2}}{2\sigma_{\tau}^{2}}}
\end{equation}
\begin{equation}\label{eq:gaussian}
    \sigma_{\tau} = \frac{\tau}{2\sqrt{2\ln(2)})}
\end{equation}
Here $\omega_{x}$ is the incident photon energy. $D^{\boldsymbol{\mu}}_{ij}$ is the transition dipole moment (TDM) matrix element between states $i$ and $j$, directed by the polarization vector $\boldsymbol{\mu}$. The electric field $\mathcal{E}(t)$ is described by a Gaussian with a peak height determined by the pulse intensity, $I_{max}$, which is given by $I_{max} = E_{max}^{2}/8\pi\alpha$, where $E_{max}$ is the maximum electric field strength, $\alpha$ is fine-structure constant, and $\tau$ is the full-width-half maximum (FWHM) pulse duration. Decay widths contribute to population losses. The ground state has a single population loss mechanism due to photoionization from the x-ray pulse, such that its decay rate, $\Gamma_{g}(t)$, is given by
\begin{equation}
    \Gamma_{g}(t) = \Gamma_{x}(t).
\end{equation}
Here, 
\begin{equation}
    \Gamma_{x}(t) = \frac{\sigma^{tot}_{\omega_{x}}}{\omega_{x}}\mathcal{E}^{2}(t),
\end{equation}
where $\sigma^{tot}_{\omega_{x}}$ is the total cross-section for photoionization at a given photon energy. We use the sum of atomic cross-sections for 2$\times$ nitrogen and 1$\times$ oxygen performed by the Hartree-Fock-Slater (HFS) method\cite{j1993atomic} to give a total cross-section of 0.169 mBarn at 400 eV. An additional contribution is included for the core-hole states due to the core-excited state lifetimes, which predominantly undergo Auger-Meitner decay, such that the decay rates for the \SCH and \DCH states can be expressed as  
\begin{equation}
    \Gamma_{s}(t) = \Gamma_{x}(t) + \Gamma^{s}_{a}
\end{equation}
\begin{equation}
    \Gamma_{d}(t) = \Gamma_{x}(t) + \Gamma^{d}_{a}
\end{equation}
In this paper, the linewidths of the \SCH and \DCH states are 0.11 eV\cite{miron2001nuclear} and 0.22 eV  respectively, where we assume the latter decay twice as fast.  Using these approximated decay rates and by neglecting coherence terms we can evaluate the evolution of the population losses by integrating over the state populations and decay rates,
\begin{equation}
   b^{a}_{n}(t) = \Gamma^{n}_{a}\int^{t}_{-\infty}dt'|c_{n}(t')|^{2}
\end{equation}
\begin{equation}
   b^{x}_{n}(t) = \int^{t}_{-\infty}dt'\Gamma^{n}_{x}(t')|c_{n}(t')|^{2}
\end{equation}
where $b^{a}_{n}(t)$ and $b^{x}_{n}(t)$ are the respective Auger-Meitner and photoionization population loss channels for state $n$. The differential EOM's to the TDSE in Eqs. \ref{eq:eomg}, \ref{eq:eoms} and \ref{eq:eomd}, were solved numerically using the 4th-order Runge-Kutta algorithm (RK4). A small time step of $5.0 \times 10^{-6}$ fs was used to minimize the population loss due to the RK4 numerical error. Initial conditions of lim$_{t\rightarrow-{\infty}}c_{(g)} =1 $, lim$_{t\rightarrow-{\infty}}c_{(s)} = 0$ and lim$_{t\rightarrow-{\infty}}c_{d} = 0$ were set,  with the peak maximum of the pulse at $t = 1$ fs. The orbital energies and transition dipole moments used in the TDSE simulations were taken from the XANES spectra RASPT2 calculations discussed in the previous section. To account for the finite bandwidth of the incident x-ray pulse, $\sigma_{x}$, we include bandwidth averaging to the quantum state population.  In this work 100 photon energies, $\omega_{j}$ in the range of
\begin{equation}\label{eq:bandwidth2}
    \omega_{n} = \omega_{x}-3\sigma_{x},..,\omega_{x},..,\omega_{x}+3\sigma_{x}.
\end{equation}
We apply a weighting factor, 
\begin{equation}\label{eq:bandwidth1}
    g\left(\omega_{n},\omega_{x}\right)= \frac{exp(\frac{-\left(\omega_{n}-\omega_{x}\right)^{2}}{2\sigma_{x}^{2}})}{\sqrt{2\pi\sigma_{x}^{2}}}
\end{equation}
to populations calculated for each of these photon energies. The TDSE was simulated in the open-source TARANTULA code\cite{fouda2022tarantula}.

\subsection{Auger-Meitner Spectra}\label{sec:augercalc}  

For the \SCH Auger-Meitner decay calculations, initial states were calculated using RASSCF(18,1,1;1,8,2) with either the N$_{t}$1s or N$_{c}$1s placed in RAS1. The two unoccupied $\pi^{*}$ orbitals were placed in RAS3. For the final cationic doublet states of the \SCH decay, core-orbitals were removed from RAS1 and 136 states were included in the SA-RASSCF(15,1,1;0,8,2) and Auger-Metiner spectra calculations. For the \DCH states, the 4 initial states of the 1st decay step were calculated by SA-RASSCF(20,2,2;2,8,2), both the N$_{t}$1s and N$_{c}$1s orbitals in RAS1 and the doubly excited state (DEXS) scheme applied. For the K$^{-1}$ final cation doublet states of the first step, 317 doublets states were considered using the active space from the \SCH initial state calculation with one less electron (SA-RASSCF(17,1,1;1,8,2)). For the second \DCH Auger-Meitner decay step, the initial states are the final states of the first step and 1627 dication singlet final states were used in the SA-RASSCF(14,1,2;0,8,2) calculation. For simplicity only the singlet dicationic final states were considered, as triplet states are expected to have a smaller contribution to the Auger-Meitner intensity\cite{tashiro2011auger}. The energies of all the states were improved by single state RASPT2 calculations with an imaginary shift of 0.4 a.u.. The configuration interaction (CI) vectors and Mulliken populations used in the determination of the rates discussed below were taken from the RASSCF wavefunctions.
 
While there have been recent advances in the calculation of Auger-Meitner decay in molecules via approaches explicitly including the continuum electron wavefunction\cite{grell2019multireference,grell2020multi}, these will become computationally too expensive for decay cascades. Therefore we adopt an approximate approach which avoids calculation of the continuum wavefunction, by estimating the two-electron integrals governing the Auger-Meitner decay by the Mulliken populations of the the valence-hole orbitals, on the wavefucntion of the core-hole initial state\cite{mitani2003theoretical,tashiro2011auger}. The approach is also applied to the prediction of the \DCH TEOE spectra and is briefly outlined here. By assuming a two-step process, where the initial excitation is decoupled from the Auger-Meitner decay, the intensities of the latter can be described by Wentzel's formula, 
 \begin{equation}
     I_{fi} = 2\pi|\langle\psi_{f}|\hat{H}-E||\psi_{i}\rangle|^{2}\equiv2\pi|t^{2}|,
 \end{equation}
$\psi_{i}$ and $\psi_{f}$ are the initial and final state wave functions, $\hat{H}$ is the Hamiltonian, $E$ is the energy of the initial state, and $t$ is the amplitude of the Auger-Meitner transition. Let's consider the decay of a core hole orbital in the initial state ($c$) that produces two valence holes in molecular orbital $v$ and $w$ and an electron in continuum wavefunction $k$. 
$t$ can be estimated by the Mulliken populations of $v$ and $w$ on the atom containing $c$ in the initial state, noted by $P_{v}(C)$ and $P_{w}(C)$ respectively
This approach is discussed in details for both resonant and non-resonant Auger-Meitner processes elsewhere\cite{mitani2003theoretical,TAKAHASHI2001137,tashiro2011auger}. The expression used in this work for multi-reference wavefunctions are shown below,
 \begin{equation}
     ^{2}t_{vv} = (cv|kv) \approx KP_{v}(C) \quad (v=w)
 \end{equation}
 \begin{equation}
     ^{2}t_{vw} = \sqrt{\frac{1}{2}}\{(cv|kw)+(cw|kv)\}  \\
     \approx \sqrt{\frac{1}{2}}K\{P_{v}(C)+P_{w}(C)\} \quad (v\neq w).
 \end{equation}
where $(cv|kv)$ are the usual two electron repulsion integrals\cite{mitani2003theoretical}. These formulas apply to the production of the doublet states in the \SCH and first step of the \DCH state decay. They can also be applied to the second step of the \DCH decay yielding singlet dicationic states. By assuming $K$ to be constant for all pairs of $v$ and $w$ \cite{mitani2003theoretical,tashiro2011auger}, the Auger-Meitner intensity between multiconfigurational initial and final state wavefunctions can be approximated by the expression below, 
\begin{equation}\label{eq:full}
    I_{n} \propto 2\pi|t_{n}|^{2}=2\pi \Big|\sum^{N_{val-occ}}_{v\leq w} C_{if,vw}t_{vw}\Big|^{2}.
\end{equation}
The $C_{if, vw}$ coefficients are determined by the products of the CI expansion coefficients between the initial and final states corresponding to the creation of holes in MOs $v$ and $w$. This simplfied approach is readily applicable to estimate multistep Auger-Meitner decay cascades in molecules and has been previously applied to double core ionized decay cascades\cite{tashiro2011auger}. In this work we apply it to the two-step decay cascade of the double core excited states, where the intensities of the second step are normalized with respect to the corresponding first step. For example, starting from the $K^{-1}K^{-1}V^{2}_{M}$ state, the intensity of the decay from the corresponding SCH cation states ($[K^{-1}V^{2}]^{+M}_{P}$) to the set of all possible valence dication states ($[K^{2}V^{2}]^{++P}_{Q}$) is labelled as ($I([K^{-1}V^{2}]^{+M}_{P},[K^{2}V^{2}]^{++P}_{Q})$) and normalized via,
\begin{equation}
    \sum_{Q}I([K^{-1}V^{2}]^{+M}_{P},[K^{2}V^{2}]^{++P}_{Q}) = I(K^{-1}K^{-1}V^{2}_{M},[K^{-1}V^{2}]^{+M}_{P}).
\end{equation}
The TEOE spectra intensity is also calculated using this approach, however the sum of the Mulliken populations on both the atomic sites is used. All Auger-Meitner spectra were computed in the open-source AUGER-POP code\cite{fouda2022augerpop}.

\subsection{Core-Electron Binding Energy Chemical Shifts}\label{sec:cebecalc}
Accurately and efficiently attaining CEBE shifts has advanced rapidly in recent years due to the work of Besley and others\cite{TAKAHATA200369,tolbatov2017benchmarking,Fouda2017,HANSONHEINE2018279,Besley2021,Zheng2020,doi:10.1021/acs.jctc.8b01071,doi:10.1080/00268976.2020.1769872,doi:10.1063/5.0013538} . In this work Attaining O and N 1s chemical shifts for valence, \SCH and \DCH states within a consistent level of theory is non-trivial as single, double and triple core-hole states are required. Fortunately  OpenMolcas RASSCF calculations provide a flexible approach to multiple core-hole states as HEXS and double excited state scheme (DEXS)\cite{delcey2019efficient} can be applied to either RAS1, RAS3 or both in combination with restricting the maximum number of electrons in RAS3. The N$_{t}$, N$_{c}$ and O 1s binding energies of the ground, valence cation and dication states were calculated by $\Delta$RASPT2(18,1,2;1,11,0), $\Delta$RASPT2(17,1,2;1,11,0) and $\Delta$RASPT2(16,1,2;1,11,0) respectively where the core orbital under question was placed into RAS1 and HEXS used to calculate core-hole with $N_{elec}-1$ active electrons in the same active space. 

The N$_{t/c}$ ss-DCH binding energies of neutral and cation N$_{t/c}$\SCH states  were calculated by $\Delta$RASPT2(20,2,2;1,11,1) and $\Delta$RASPT2(19,2,2;1,11,1) respectively. The initial state was generated by using HEXS on the desired N$_{t/c}$ core hole in RAS1 (the other N$_{t/c}$ was in RAS3). The ss-DCH binding energy was calculated by applying DEXS to RAS1 of the N-1 wavefunction. To calculate ts-DCH binding energies, HEXS was applied to both the RAS1 and RAS3 subspaces. This same approach was applied to the O1s ts-DCH binding energies.

In order to calculate the binding energies of the N$_{t/c}$\DCH state a triple core-hole final state is required. The initial state was calculated at the RASPT2(20,2,2;1,11,1) where the HEXS is applied to both RAS1 and RAS3 each containing either the N$_{t}$ or N$_{c}$ orbitals. The binding energy is calculated by applying both DEXS to RAS1 and reducing the maximum number of electrons in RAS3 to 1. This yields a RASPT2(19,2,1;1,11,1) calculation. To calculate the O1s binding energy, the initial state was prepared by a RASPT2(22,2,2;2,11,1) calculation with both N$_{t}$ and N$_{c}$ 1s orbitals in RAS1 with DEXS applied. The O1s orbital was in RAS3 and the maximum number of electrons in RAS3 was changed from 2 to 1 like the N$_{t/c}$ binding energy calculations.

\section{Results and Discussion}\label{sec:results}
The present work studies the formation of \DCH states during resonant interaction with an intense few-to-sub femtosecond x-ray pulse. N$_{2}$O represents a model system containing two distinct atomic centers of the same element (N) with a significant chemical shift of about 4 eV. In following we present the calculated RASPT2 energetics of the two-step XANES process in Subsection \ref{sec:xanes}, TDSE simulations of the production of the \DCH states using an ultrashort intense pulses in Subsection \ref{sec:tdse}. Subsections \ref{sec:auger} and \ref{sec:cebe} look at the characteristic Auger-Meitner decay spectra and core electron binding energies of the \SCH and \DCH states respectively. 

\subsection{XANES Spectra}\label{sec:xanes}
\begin{figure*}[h]
\centering
\includegraphics[height=8.55cm]{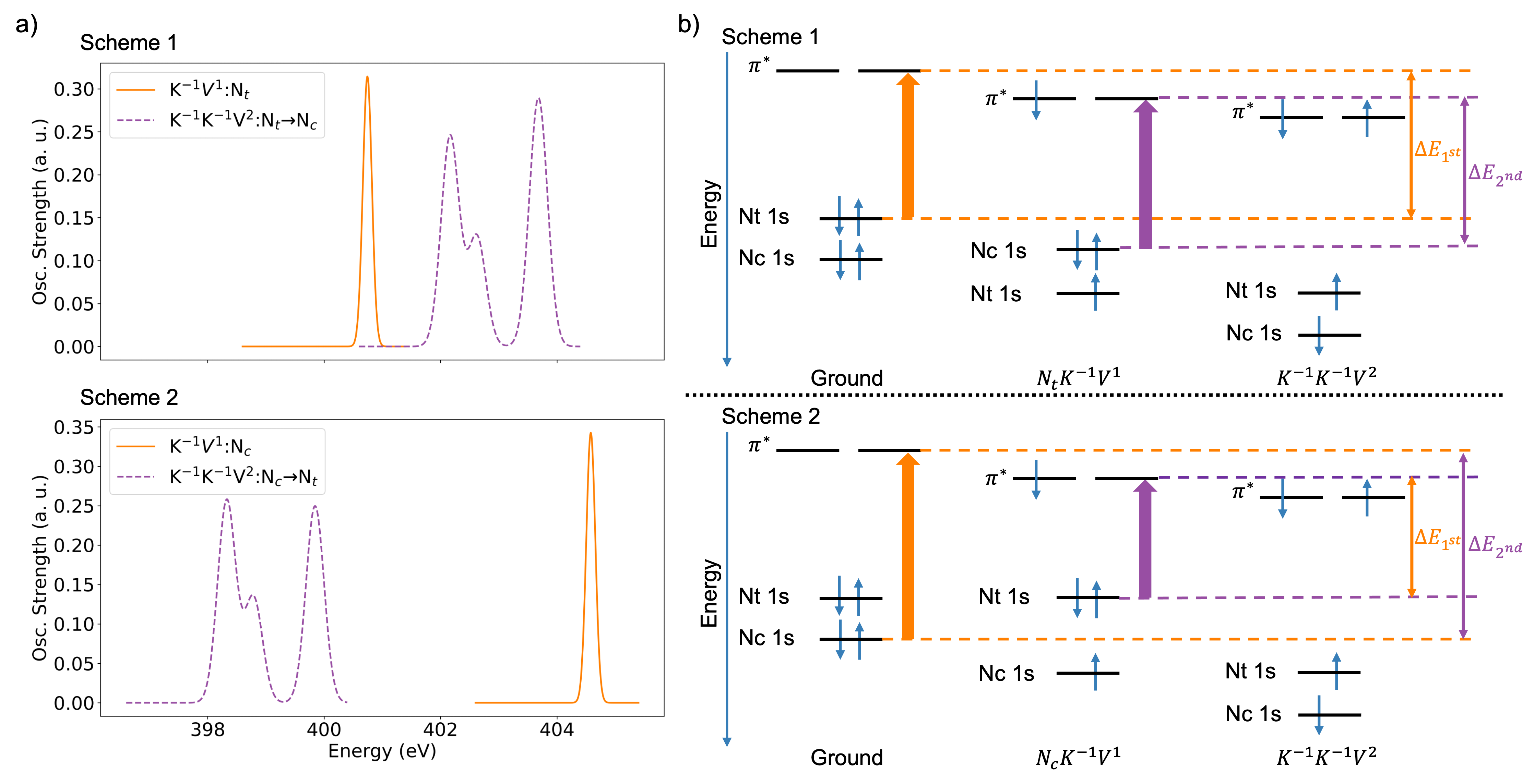}
\caption{a) RASPT2 calculated XANES spectra for producing \SCH and \DCH states through sequential core-excitations in N$_{2}$O at the nitrogen K-edge to the doubly degenerate $\pi^{*}$ orbital. Top Panel: Scheme 1 involving K-shell excitation at N$_{t}$ then N$_{c}$. Lower Panel: Scheme 2 involving K-shell excitation at N$_{c}$ then N$_{t}$. b) Energy level diagrams (not to scale), comparing double core-excitation Schemes 1 and 2. Using the $^{1}\Sigma^{-}$ \DCH state values in Table \ref{tbl:xas}, Scheme 1 |$\Delta E_{1}$-$\Delta E_{2}$| = 1.88 and Scheme 2 |$\Delta E_{1}$-$\Delta E_{2}$| = 5.79.}
\label{fgr:xas}
\end{figure*}

 The corresponding N1s to $\pi^{*}$ core excitation energies differ by 3.7 eV\cite{larkins1988core} and a strong discrimination between the sites can be expected for the first excitation step with sub-to-few femtosecond pulse bandwidths. RASPT2 calculated XANES spectra of the double core excitation of N$_{2}$O  are shown in Fig \ref{fgr:xas} a). The orange and purple peaks correspond to the first and second core-excitation's respectively. The top and bottom panels show two possible excitation schemes that differ by the order of the N$_{c}$ and N$_{t}$ K-shell excitation. Scheme 1 excites at the N$_{t}$ then N$_{c}$ 1s sites and Scheme 2 excites at the N$_{c}$ then N$_{t}$ 1s sites. The energies for all the transitions are shown in Table \ref{tbl:xas} and the single core-excitation energies from the ground state are shown to closely match  experimental values. 

The second core-excitation step involves three products, as it involves two electrons shared across two degenerate $\pi^{*}$ orbitals, analogous to single core-excitations in NO\cite{fouda2020observation,fouda2021site}. The symmetry labels and the excitation energies from the single core excited states are shown in Table \ref{tbl:xas}. The predicted values for the second excitation's to the \DCH states rely on the close agreement of the theoretical and experimental \SCH results and are supported by previous studies where RASPT2 agrees well with experimental DCH state energies\cite{delcey2019efficient,koulentianos2020high}. While it is well established that the binding energy of core electrons increases significantly when the molecule is core ionized once, we predict that there is shift in the core-excitation energy of both the N$_{c}$ and N$_{t}$ atoms to the $\pi^{*}$ orbital of approximately 2 eV towards lower binding energy following core-excitation at the nitrogen site to the same $\pi^{*}$ orbital compared to the ground state values. This can be understood as the result of the stabilization (lower energy) of valence shells due to a higher charged core, while the CEBE remains essentially the same in the still neutral molecule (except for the second electron at the same site of the first core-excitation). This effect can be quantified by looking at the RASSCF canonical orbital energies shifts, between the ground and \SCH states. In the N$_{t}$ \SCH state, the N$_{c}$ 1s orbital energy shifts by +0.37 eV and the $\pi^{*}$ shifts by -5.46 eV and in the N$_{c}$ \SCH state, the N$_{t}$ 1s orbital energy shifts by -0.87 eV and the $\pi^{*}$ shifts by -6.74 eV. This effect is demonstrated by the spectra and energy level diagrams in Fig. \ref{fgr:xas} a) and b) respectively.
 
To enable sequential core-excitations N$_2$O, Fig \ref{fgr:xas} a) and b) shows that a pulse centered at 401 eV with a bandwidth of around 2 eV should selectivity excite the N$_{t}$ ($\approx$401 eV) then N$_{c}$ ($\approx$402 eV) K-shell electrons to the $\pi^{*}$. The alternative double core-excitation scheme (N$_{c}$ then N$_{t}$) requires a first photon at $\approx$405 eV and a second photon at $\approx$398 eV and is much less probable.

 \begin{table}[h]
\small
  \caption{RASPT2 calculated excitation energies for producing \SCH and \DCH states through sequential core-excitations in N$_{2}$O at the nitrogen K-edge to the doubly degenerate $\pi^{*}$ orbital. Experimental \SCH taken values from ref\cite{larkins1988core}. Scheme 1 and 2 correspond to the top and bottom panels in Fig \ref{fgr:xas} a) respectively.}
  \label{tbl:xas}
  \begin{tabular*}{0.48\textwidth}{@{\extracolsep{\fill}}lll}
    \hline
    Transition & $\Delta$RASPT2 & Experiment \\
    \hline
    Scheme 1\\
    $K^{-1}V^{1}$:N$_{t}$& 400.74 & 401.13  \\
    $K^{-1}K^{-1}V^{2}:N_{t}\rightarrow N_{c}^{1}\Delta$     & 402.17 & - \\
    $K^{-1}K^{-1}V^{2}:N_{t}\rightarrow N_{c}^{1}\Sigma^{-}$ & 402.62 & - \\
    $K^{-1}K^{-1}V^{2}:N_{t}\rightarrow N_{c}^{1}\Sigma^{+}$ & 403.68 & - \\
    Scheme 2\\
    $K^{-1}V^{1}:N_{c}$ & 404.58 & 404.73  \\
    $K^{-1}K^{-1}V^{2}:N_{c}\rightarrow N_{t}^{1}\Delta$    & 398.33 & - \\
    $K^{-1}K^{-1}V^{2}:N_{c}\rightarrow N_{t}^{1}\Sigma^{-}$& 398.79 & - \\
    $K^{-1}K^{-1}V^{2}:N_{c}\rightarrow N_{t}^{1}\Sigma^{+}$& 399.85 & - \\
    \hline
  \end{tabular*}
\end{table}

\subsection{Electronic Population Transfer Dynamics}\label{sec:tdse}

\begin{figure*}[ht]
\centering
  \includegraphics[width=15.0cm]{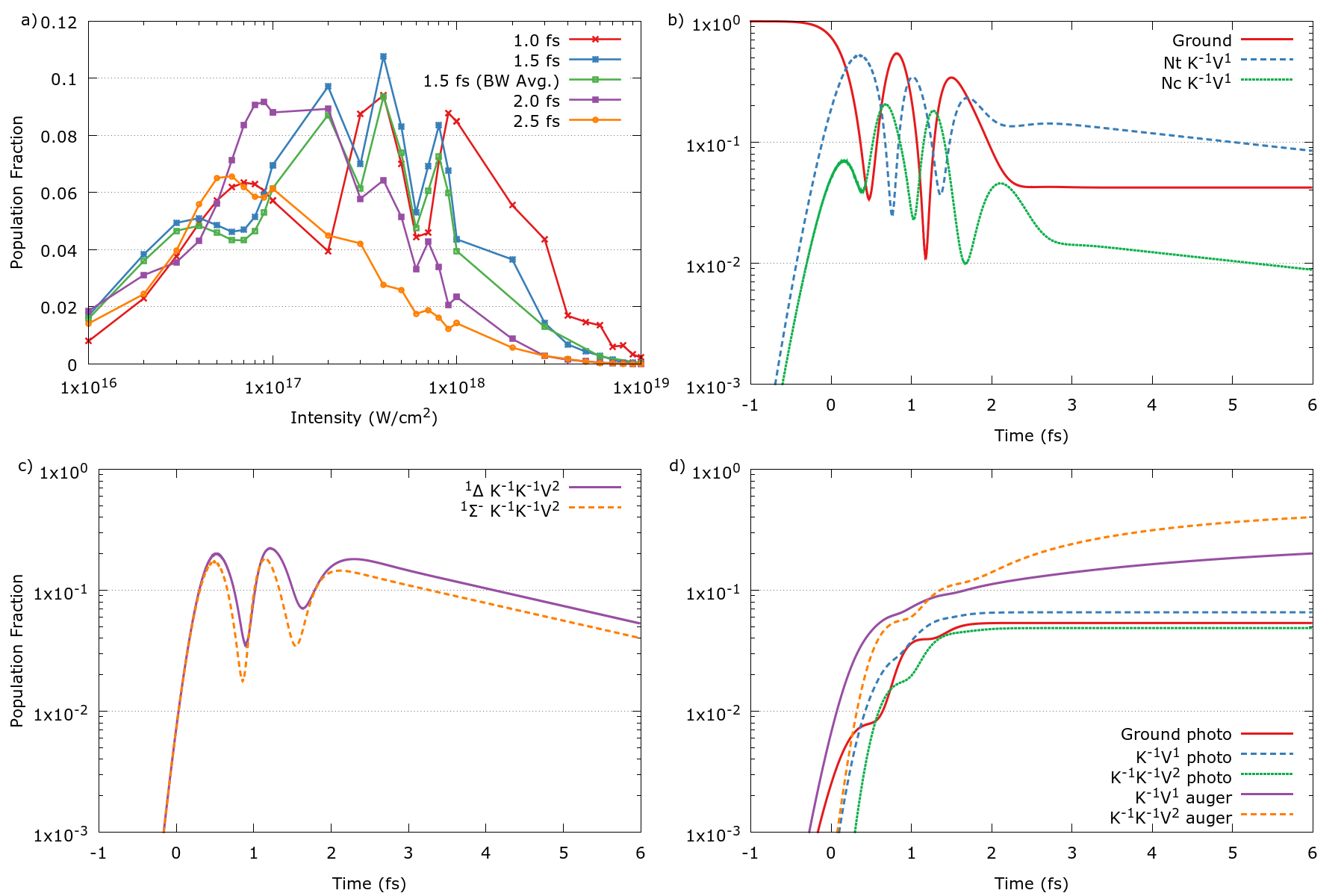}
  \caption{TDSE simulations producing \DCH states through sequential core-excitation's in N$_{2}$O at the nitrogen K-edge to the doubly degenerate $\pi^{*}$ orbital. The simulations are based on matrix elements from on MS-RASPT2 wavefunctions. In all subplots the pulse is centered at 1.0 fs and the central photon energy is 401 eV. a) The total \DCH state population at 6 fs as a function of the pulse duration and intensity. Bandwidth averaging (BW Avg.) is applied to 1.5 fs. In b-d) the pulse duration and intensity are 1.5 fs and 4$\times$10$^{17}$W/cm$^{2}$ and the results are averaged over a 1.2 eV bandwidth. b) Time evolution of the ground and \SCH state populations. c) Time evolution of the \DCH state populations. d) Time evolution of the population loss channels. }
  \label{fgr:tdse}
\end{figure*}

We examine how the pulse conditions proposed by the previous subsection can produce \DCH states via a sequential core-excitation mechanism using TDSE simulations described in Section \ref{sec:tdsecalc}. First we discuss the time and intensity scan of the incident pulse to determine optimal conditions for producing a high yield of the  \DCH states. Fig \ref{fgr:tdse} (a) shows the total \DCH population at 6 fs with a pulse centered at 1 fs with durations of 1.0, 1.5, 2.0 and 2.5 fs.  For each pulse duration, we sampled across 28 intensity points between $1\times10^{16}$ and $1\times10^{19}$ W/cm$^{2}$. The sampled high-intensity range shows a sensitivity to the total \DCH population, as seen by the multiple peaks in the population across the intensity range. This is due to the complex interplay of Rabi-cycling and pulse duration as the results are not averaged over the focal volume \cite{fouda2021site}. The maximum \DCH population in Fig \ref{fgr:tdse} a) is approximately 10$\%$ of the total population and is generated by a 1.5 fs pulse with $4\times10^{17}$ W/cm$^{2}$ intensity. This is close to the intensity of the two-photon cross-section saturation fluence of the ground to N$_{t}$ \SCH and N$_{t}$ \SCH to $^{1}\Delta$ \DCH excitation's at 5.11$\times$10$^{17}$ W/cm$^{2}$. The results show that the optimal conditions for the \DCH production are for pulse duration's between 1 and 2 fs and intensities between $1\times10^{16}$ and $1\times10^{17}$ W/cm$^{2}$. At higher intensities, the \DCH population will be reduced due to the valence photoionization. On the other hand, lower intensities will not sufficiently drive the two-photon process.


For pulse durations longer than 2.5 fs, the production of \DCH will compete with the decay of intermediate \SCH states. Furthermore, a sufficiently large bandwidth is required to first excite the N$_{t}$ \SCH state at ~401 eV then the \DCH states at ~402-403 eV. In particular, pulses with a duration of the 1-2 fs will provide a bandwidth of approximately 1-2 eV to drive this two-photon transition. Fig. \ref{fgr:tdse} (a) show the effect of pulse bandwidth for the pulse duration of 1.5 fs, in which we sample across a 1.2 eV FWHM bandwidth with a central photon energy of 401 eV. The bandwidth causes a small drop in the population transfer to the \DCH states, which may result from the increased detuning from resonance, but maintains the same overall the trend and sensitivity with respect to the intensity between $1\times10^{17}$ and $1\times10^{18}$ W/cm$^{2}$.

\begin{figure*}[ht!]
 \centering
 \includegraphics[width=15.0cm]{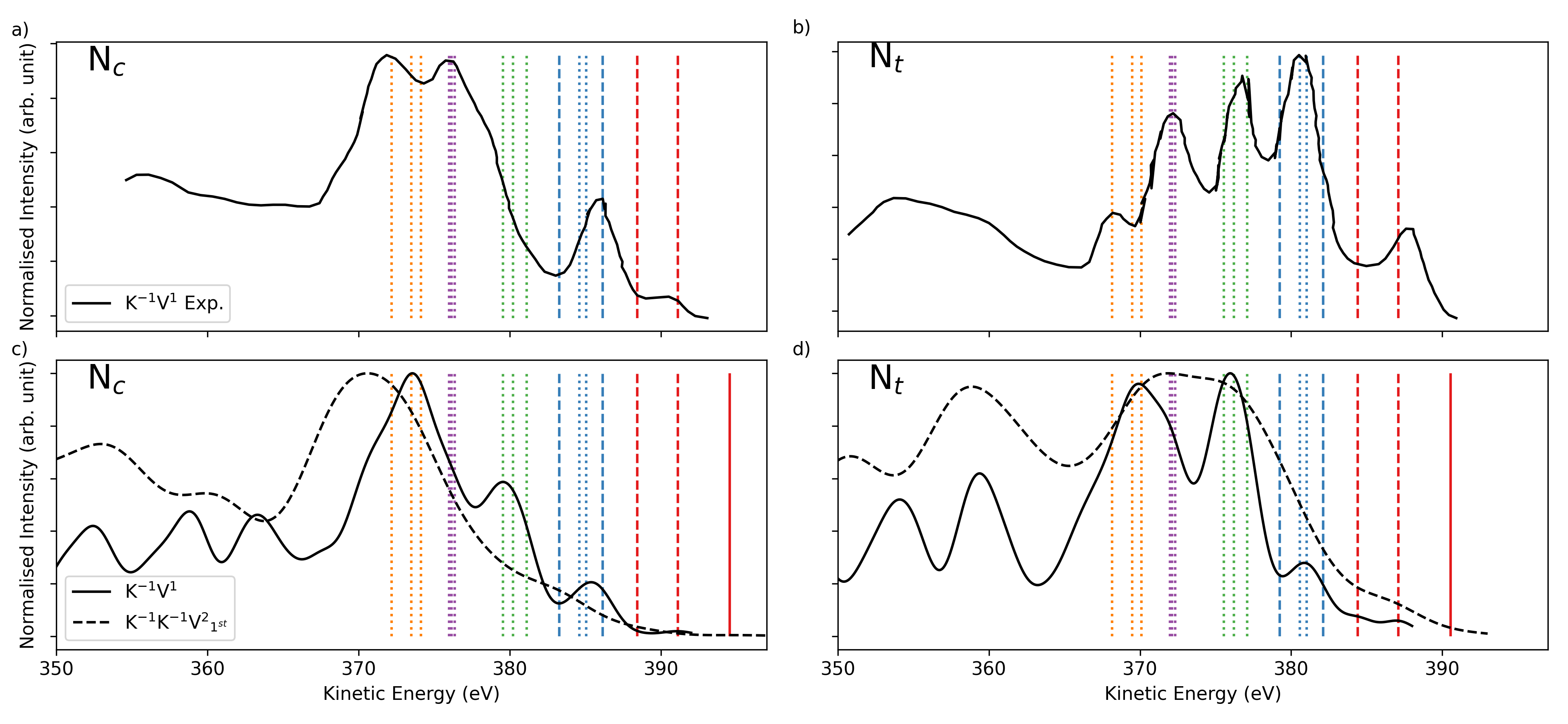}
 \caption{Analysis of the states contributing to peaks in the calculated Auger-Meitner spectra. Left (a) and c)) and right (b) and d)) columns show Auger-Meitner processes filling the N$_{c}$ and N$_{t}$ 1s core holes respectively, where a) and b) show the experimental \SCH spectra\cite{larkins1988core} and c) and d) show the calculated \SCH and  K$^{-1}$K$^{-1}$V$^{2}_{1^{st}}$ spectra. Vertical lines indicate the calculated energies of the final states. Solid: double participator transition unique to \DCH states. Dashed: participator channels. Dotted: spectator channels. Red: $\pi^{*}$-$\pi^{*}$, $\pi^{*}$-2$\pi$, $\pi^{*}$-7$\sigma$. Blue: $\pi^{*}$-1$\pi$, 2$\pi$-2$\pi$, $\pi^{*}$-6$\sigma$. Green: 2$\pi$-1$\pi$, 2$\pi$-7$\sigma$. Purple: 2$\pi$-6$\sigma$, 7$\sigma$-6$\sigma$, Orange: 7$\sigma$-6$\sigma$, 7$\sigma$-1$\pi$, 1$\pi$-1$\pi$. 3.5 eV FWHM Gaussian broadening is applied to the \SCH spectra and 7 eV FWHM Gaussian broadening is applied to the \DCH spectra.}  
 \label{fgr:auger_state}
\end{figure*}
Fig. \ref{fgr:tdse} (b) to (d) show the population transfer dynamics for producing \DCH states under an intensity of $4\times10^{17}$ W/cm$^{2}$ and 1.5 fs pusle duration. The simulation includes same bandwidth averaging applied in (a) but does not include focal volume averaging. The results are averaged over linear $x$ and $y$ pulse polarizations and the molecular bond lies in the $z$ axis. Fig. \ref{fgr:tdse} (b) shows the time-evolution of the ground and \SCH state populations, Fig. \ref{fgr:tdse} (c) shows the \DCH state evolution and Fig. \ref{fgr:tdse} (d) shows the sums of the Auger-Meitner and photoionization population losses. The decay widths of the \SCH and \DCH states are 0.11 and 0.22 eV respectively, this can be observed by the steeper gradient of the \DCH state population decline. The \DCH decay width is based on the assumption that the these states will decay twice as fast. Despite this, the yield of \DCH states is comparable to the \SCH states after the Rabi-cycling period. Fig. \ref{fgr:tdse} (d) shows that Auger-Meitner decay is the dominating loss mechanism in which the \DCH contribute to more. The first step of the \DCH decay leads to SCH cation states, these are also produced by the valence photoionization of the \SCH states, which is shown to be a smaller population loss channel in Fig. \ref{fgr:tdse} (d) . In Subsection \ref{sec:cebe} we show that these states have ss-DCH ionization values that are resolvable within 100's attosecond TR-XPS probe bandwidths, however this will not provide a direct observation of the \DCH states due to the competing photoionization of the \SCH intermediates. 

Fig. \ref{fgr:tdse} (c) shows the production of a superposition of the \DCH states including the $^{1}\Sigma^{-}$ state and the $^{1}\Delta$, which is the sum of population of $^{1}\Delta_{x}$ and $^{1}\Delta_{y}$ states. The $^{1}\Sigma^{-}$ states are not produced as a result of the transition dipole phase effect\cite{yuan2019role}. Since the transition dipole moment phases are not uniquely defined, this leads to the difference in the  $^{1}\Delta_{x}$ and $^{1}\Delta_{y}$ populations despite the results being averaged over $x$ and $y$ linearly polarized pulses. During the lifetime of the \DCH state, there will be a charge oscillation between the electron configurations of the doubly occupied $\pi^{*}$ configurations. It was previously demonstrated that an circularly polarized IR streaking field can detect the charge oscillation of a \SCH core excited superposition in NO molecules\cite{li2022attosecond}. This streaking detection technique may be applicable to probe the charge oscillation in \DCH  of N$_{2}$O.

\subsection{Auger-Meitner Spectra}\label{sec:auger}

This subsection investigates the possibility of observing the \DCH states through Auger-Meitner spectroscopy. Following resonant excitation, the decay of N$_{2}$O decays can produce electrons with different kinetic energies depending on the excitation site\cite{larkins1988core,piancastelli2007high}. Fig. \ref{fgr:auger_state} presents Auger-Meitner decay spectra, where the electron energies were determined at the RASPT2 level of theory and the intensities derived from the CI coefficients and Mulliken populations were calculated using the RASSCF wavefunctions, as discussed in Section \ref{sec:augercalc}. Due to the large number of possible final states for the \DCH decay, we do not consider the shake-off effects in this work, however, we note that shakeup states, which can play an role in DCH spectroscop\cite{doi:10.1021/acs.jpclett.0c01167}, will be considered in future work. To check the validity of our approach,  which omits the continuum electron wavefunction, we compare our simulated \SCH spectra to experiment in Fig. \ref{fgr:auger_state} and show that they are in qualitative agreement. This is further supported by the analysis of the the final states, which shows that the peaks in calculated spectrum form a one-to-one mapping with the experimental peaks. A clear signature of the calculated and experimental \SCH  decay spectra around 385 eV in the Fig. \ref{fgr:auger_state} comes from participator channels. However the calculations indicate this only applies to participator peaks involving the 2$\pi$ and 1$\pi$ orbitals, as the 7$\sigma$ and 6$\sigma$ transitions (shown by the lower energy red-dashed lines in a) and b)) lie between the peaks in the theory and experiment. Unique to the \DCH states is the double-participator decay.  Its signature is having higher kinetic energy than other pathways, as shown by the solid red lines in Fig. \ref{fgr:auger_state} at 394.5 and 390.5 eV for N$_{c}$ and N$_{t}$ respectively while the highest energy participator peaks for the \SCH decay are at 389.7 and 386.3 eV respectively. However, our calculations indicate that the signal from the double participator peak is small and the dominant signal comes from other pathways.  Future calculations with higher levels of theory\cite{grell2020multi} will further verify this finding. 

\begin{figure}[ht!]
 \centering
 \includegraphics[height=10.0cm]{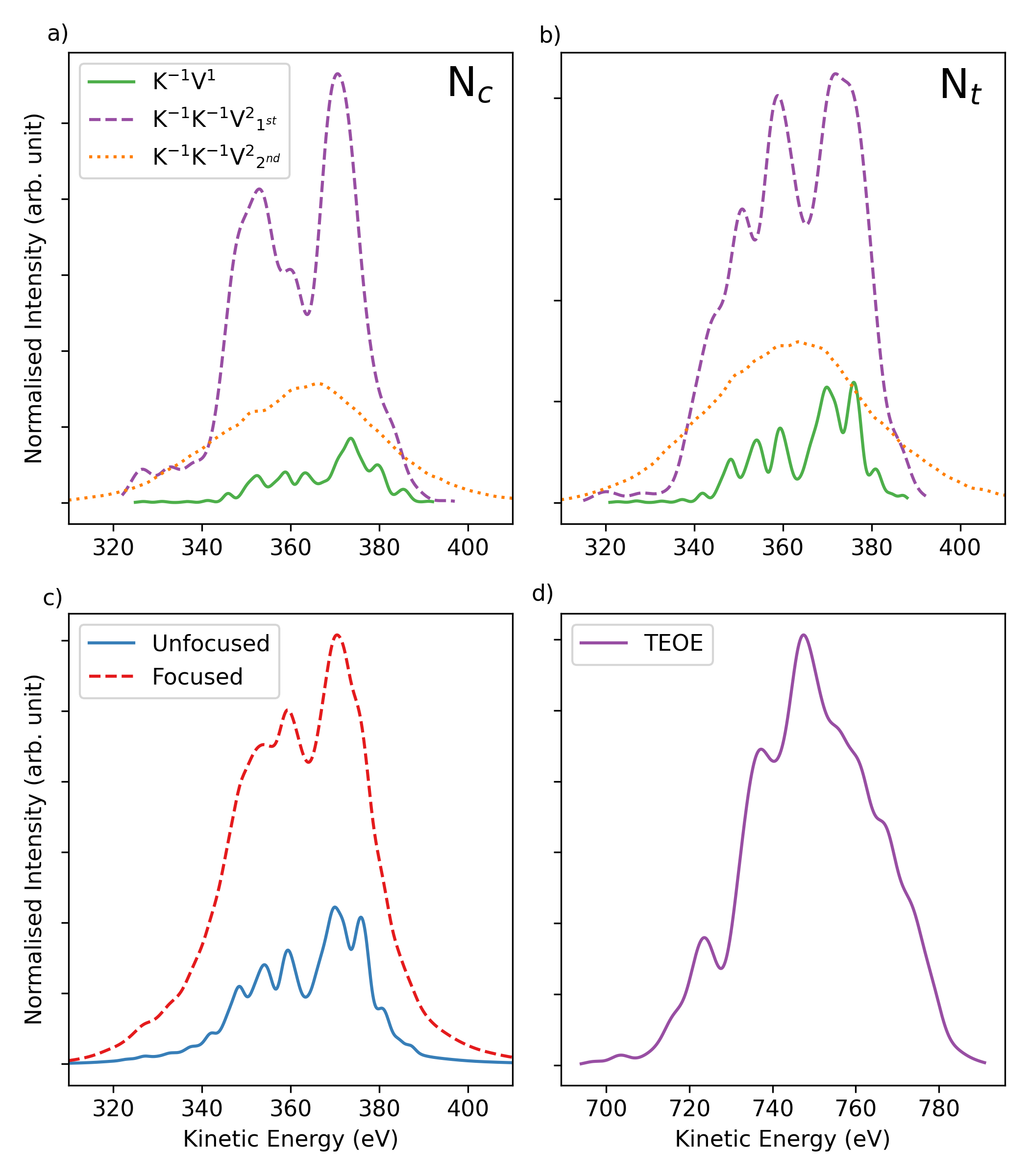}
 \caption{Comparison of the \SCH against the \DCH first and second (\DCH$_{1^{st}}$ and \DCH$_{2^{nd}}$) steps of the Auger-Meitner decay processes calculated with RASSSCF/PT2 wavefunctions and the Mulliken population analysis method to estimate the intensity. a) and b) Auger-Meitner processes filling the N$_{c}$ and N$_{t}$ 1s core holes respectively, a 0.5 weighting factor was applied to the $K^{-1}K^{-1}V^{2}_{1st}$ and $K^{-1}K^{-1}V^{2}_{2nd}$ spectra as there are 4 initial \DCH states as opposed to the two initial states for the \SCH decay. c) Estimated change in Auger-Meitner line-shape between an unfocused and focused beam. Both line shapes contain all the spectra in a) and b) and the sum of the each contributing spectra are normalized to the corresponding Auger-Meitner populations extracted from the TDSE simulation in Fig. \ref{fgr:tdse} b, c, d) at 10 fs (close to the maximum Auger-Meitner yield). Intensities of $1\times10^{16}$W/cm$^{2}$ and $4\times10^{17}$W/cm$^{2}$ (representing linear and non-linear interaction regimes) were used for the unfocused and focused lineshapes respectively. d) TEOE spectra (Fig \ref{fgr:schematic} d)), as both core-holes are filled simultaneously, the sum of the Mulliken populations on both the N$_{c}$ and N$_{t}$ atoms are used. 3.5 and 7 eV FWHM Gaussian broadening was applied to the \SCH and \DCH (incl. TEOE) spectra respectively.}
 \label{fgr:auger}
\end{figure}

The qualitative agreement shown in Fig. \ref{fgr:auger_state} enables our calculations to provide a reasonable estimation of the two-step \DCH decay cascade. Fig. \ref{fgr:auger} a) and b) show the decay processes where the N$_{c}$ and N$_{t}$ core holes are filled respectively. The first step involves the 4  neutral singlet \DCH DCH states each decaying to 316 doublet cationic $[K^{-1}V^{1}]^{+}$ single core-hole (SCH) states. These states will then undergo a second decay step to either singlet or triplet dicationic valence states. In our calculation we have included 1627 potential final singlet dicationic states and exclude the triplet states to reduce the cost of the calculation, which are expected to have a smaller contribution to the Auger-Meitner intensity\cite{tashiro2011auger}. The \SCH decay has two initial states as either the $\pi^{*}_{x}$ or $\pi^{*}_{y}$ orbital can be singly occupied and there are 4 initial states for the first step \DCH decay, therefore both steps of \DCH decay have been weighted by a factor of 0.5.


By using a central photon energy of 401 eV, there is an increased selectivity for starting the sequential core-excitations via a core-hole in N$_{t}$, according to the energetics found in Fig. \ref{fgr:xas} and Table \ref{tbl:xas}. The shape of the Auger-Meitner spectrum should therefore be very sensitive to the intensity of the x-ray pulse as the additional channels involving  N$_{c}$ core hole decay will only appear when the second core-excitation occurs to form the \DCH state superposition. In Fig. \ref{fgr:auger} c), we predict the change in spectral shape with respect to the intensity.  Here, the Auger-Meitner spectra from an unfocused (low-intensity) beam, is compared to the case of a focused (high-intensity) beam, which are determined by combining the spectra in Fig. \ref{fgr:auger} a) and b), where the sum of the each contributing spectra are normalized to the corresponding Auger-Meitner populations extracted from the TDSE simulation in Fig. \ref{fgr:tdse} b, c, d) at 10 fs (close to the maximum Auger-Meitner yield). Intensities of 1$\times$10$^{16}$W/cm$^{2}$ and 4$\times$10$^{17}$W/cm$^{2}$ (representing linear and non-linear interaction regimes) were used for the unfocused and focused lineshapes respectively.

Fig. \ref{fgr:auger} a) and b) illustrates that the second step of the \DCH decay results in the creation of a broad background signal, which is not resolvable, but should not hinder observation of the first steps due to the relatively low intensity which is dependent on the intensities of the first step. In addition, since the Auger-Meitner spectra often fall on top of valence photoemission lines, it is important to further analyze the resonant features either from angular distribution of the emitted electrons \cite{knie2014angle} or by scanning the incident photon energy over a small range. Throughout this work we have considered a fixed nuclear geometry and expect relatively little nuclear motion during the core-hole lifetimes. However it has been observed experimentally that Auger-Meitner spectroscopy can be sensitive to nuclear motion within the lifetime of the core-excited\cite{hjelte2001evidence} and previous studies have shown that the \SCH states in N$_{2}$O adopt a bent geometry\cite{lebrun1993dissociation}.

Furthermore, we examine a TEOE process (see the schematic in Fig. \ref{fgr:schematic} d)) that involves the simultaneous collapse of two electrons to fill both core vacancies in the \DCH states and the emission of a single electron to form valence cation states . Our calculated spectrum in Fig. \ref{fgr:auger} d) shows that TEOE results in a large energy transfer to the single continuum electron, yielding significantly higher kinetic energies (700-800 eV) compared to the energies of other Auger-Meitner electrons, if this energy range is experimentally detectable then the TEOE electrons will provide a clear signature of the \DCH state observation.

\subsection{Core Electron Binding Energy Chemical Shifts}\label{sec:cebe}
\begin{table*}
\small
  \caption{RASPT2 calculated CEBE values in eV for the products of the sequential core excitation and subsequent Auger-Meitner decay processes. The values represent TR-XPS values at the O K-edge and N K-edge. Due to the nature of the \SCH and \DCH states the N K-edge values consider SCH and DCH states. The DCH state probes can either be single-site (ss) or two-site (ts). Only the CEBE's of the lowest spin ground state configurations are shown for simplicity.}
  \label{tbl:cebe}
  \begin{tabular*}{\textwidth}{@{\extracolsep{\fill}}llllllll}
    \hline
    Species                & O K-edge & Nt K-edge SCH & Nc K-edge SCH & Nt K-edge ss-DCH & Nt K-edge ts-DCH & Nc K-edge ss-DCH & Nc K-edge ts-DCH \\
    \hline
    Neutral\\
    G                 & 541.41   & 412.35       & 408.25       & -                & -                & -                & -\\
    $K^{-1}V^{1}$:N$_{t}$    & 540.58  & -             & -             & 475.80           & -                & -                & 407.77\\
    $K^{-1}V^{1}$:N$_{c}$    & 541.30  & -             & -             & -                & 411.52           & 479.99           & - \\
    $K^{-1}K^{-1}V^{2}$(avg.)& 541.37  & -             & -             & 476.11           & -                & 476.11           & -\\
    Cation\\
    $[K^{2}V^{1}]^{+}$            & 554.618  & 422.34        & 420.59        & -                & -                & -                & -\\
    $[K^{-1}V^{1}]^{+}:N_{t}$   & 552.45  & -             & -             & 487.75           & -                & -                & 420.39\\
    $[K^{-1}V^{1}]^{+}:N_{c}$   & 553.29  & -             & -             & -                & 424.51           & 492.01           & -\\
    Dication\\
    $[K^{2}V]^{++}$           & 566.12  & 431.91        & 432.14        & -                & -                & -                & -\\
    \hline
  \end{tabular*}
\end{table*}
While the deexcitation spectrum should provide a clear signature of the production of neutral excited ts-DCH states, it is unlikely to be sensitive to the dynamics of these states after their formation. To further establish their existence and gain a first glance of the complex decay cascade, we predict chemical shift values of these excited states which can be measured with TR-XPS. In practise, an attosecond pulse would be necessary to resolve the decay steps with lifetimes between 3 and 5 fs. Therefore, chemical shifts must approximately exceed 5 eV to be resolvable within 100's attosecond pulse bandwidths, even if techniques such as spectral ghost imaging (spooktroscopy) present some promising avenues\cite{driver2020attosecond} for high energy resolution with attosecond x-rays. Table \ref{tbl:cebe} shows the predictions for the chemical shifts at the three atomic sites between the neutral ground, \SCH and \DCH states and the associated intermediate and final states of the Auger-Meitner decay, calculated by $\Delta$RASPT2 as described in Subsection \ref{sec:cebecalc}. In cases where the removal of an electron yields multiple spin states, only the low-spin cases are presented as the spin multiplicity has a minor effect on the chemical shifts. Also, the values in Table \ref{tbl:cebe} represent lower bounds to the chemical shifts as they only consider the ground state of each particular low spin species in the core-excitation and Auger-Meitner decay. 

Between the neutrally charged species, less than 1 eV chemical shifts in the O K-edge binding energies were predicted around 541 eV, which likely prevents them from being resolved within the hundreds of attoseconds pulse bandwidths currently available at XFEL light sources. Furthermore, there is an overlap between the valence cation final states ($[K^{2}V^{1}]^{+}$) from the \SCH Auger-Meitner decay and the single-core-hole cation intermediate ($[K^{1}V^{2}]^{+}$) states from the first step in the \DCH state decay of about 2 eV around 553 eV. Unresolvable chemical shifts within hundreds of attoseconds pulse bandwidths also occur for the single, two-site (ts) double and single-site double CEBE's for the neutral species at the N K-edge. On the other hand, our calculations indicate that probing the nitrogen sites through the ionization of half-empty orbitals is more promising. Specifically, the binding energy of a ss-DCH at either the central or terminal nitrogen site is different in the singly core-excited cationic states produced following the first decay step of the \DCH state decay when compared with the singly or doubly excited neutral molecules. The 10 to 20 eV shift is easily resolvable even with the large bandwidth of a 100's of attosecond probe pulse. These transient states will have a short ($\approx$3 fs) lifetime but the TDSE simulations presented in Fig. \ref{fgr:tdse} show that the population produced by the \DCH decay is significantly higher than the population produced by valence photoionization of the \SCH states by the pulse. By changing the delay between the pump and probe pulses, one could probe the dissociation and fragmentation of the di-cationic states produced from Auger-Meitner decay.

\section{Conclusions}

We have simulated the production of \DCH states by selective sequential double core excitations at the N K-edge of N$_{2}$O to the doubly degenerate $\pi^{*}$ orbitals and have shown a number of possible avenues for detecting these states by exploiting the contemporary capabilities of XFEL facilities. The simulations are based on RASSCF multiconfigurational quantum chemical calculations which offer accurate prediction of multiple hole states in gas-phase small molecules. The XANES spectra of the sequential double core-excitation was predicted using RASPT2 calculations, showing excellent agreement to experiment for the \SCH excitation energy and providing a reliable estimate for the energetics of the \DCH state production. We found that the second core-excitation energy from the adjacent, unexcited nitrogen atom is approximately 2 eV lower that its original ground state excitation energy. As a result, a 1-fs pulse centered at 401 eV with a 2-eV bandwidth would enable the production of neutral doubly core-excited states preferably via a sequence of N$_{t}$ 1s excitation followed by N$_{c}$ 1s excitation and suppress the opposite sequence. This is supported by the TDSE simulations that demonstrate a femtosecond pulse produces a high yield of the \DCH states and a superposition of the $^{1}\Delta$ and $^{1}\Sigma$ states.

We show that by examining the pulse intensity dependence of the site-selective Auger-Meitner decay spectra, the change of spectra lineshape can reveal a signature of the \DCH states. Furthermore the high kinetic energies ($\approx$700-800 eV) of the TEOE electrons will provide an additional direct signature of the \DCH states. The RASPT2 calculations show that most of core ionization energies of the core-excited states and decay intermediates and final states overlap within the hundreds of attosecond pulse bandwidths ($\approx$5 eV) and therefore are not resolvable. The exception is the ss-DCH ionization energies of the single core-excited cation states, in which the CEBE differ by more than 5 eV with respect to other states. Also, our TDSE simulations demonstrate that Auger-Meitner decay of the \DCH states is the dominant pathway for producing single-core-hole cation states in comparison to the pathway of the valence photoionization of the \SCH states.

We have produced a clear theoretical demonstration of the potential to produce and detect these states in ultrafast, intense XFEL experiments and there experimental validation would provide important insights into the presence of sequential resonant interactions for future experiments at the forefront of the inducing and tracking molecule dynamics in these pulse conditions. 

\section*{Conflicts of interest}

\section*{Acknowledgements}
This material is based on work supported by the U.S. Department of Energy, Office of Basic Energy Sciences, Division of Chemical Sciences, Geosciences, and Biosciences through Argonne National Laboratory. Argonne is a U.S. Department of Energy laboratory managed by UChicago Argonne, LLC, under contract DE-AC02-06CH11357.



\balance


\bibliography{rsc} 
\bibliographystyle{rsc} 

\end{document}